\documentclass[12pt]{article}
\usepackage{graphicx}
\usepackage{bm}
\begin{document}

\textwidth 16cm
\textheight 24cm
\hoffset -1.5cm
\baselineskip 20pt

\centerline{Baryonic Effect on $\chi_{cJ}$ Suppression in Au+Au Collisions at 
RHIC Energies
\footnote{Supported by National Natural Science Foundation of China under 
Grant No. 10075022, and X.-M. Xu supported in part by National Natural Science 
Foundation of China under Grant No. 10135030, in part by the CAS Knowledge 
Innovation Project No. KJCX2-SW-NO2.}}
\centerline{Ru Peng$^1$, Xiao-Ming Xu$^{2,3}$, Dai-Cui Zhou$^1$}
\centerline{$^1$ Institute of Particle Physics, 
Central China Normal University,}
\centerline{Wuhan 430079, China}
\centerline{$^2$ Department of Physics, Shanghai University, Baoshan, 
Shanghai 200444, China}
\centerline{$^3$ Shanghai Institute of Nuclear Research, Chinese Academy of 
Sciences}
\centerline{PO Box 800204, Shanghai 201800, China}
\date{}

\begin{abstract}
We predict that initially produced $\chi_{cJ}$ mesons at low transverse 
momentum in the central rapidity region are almost dissociated by nucleons and 
antinucleons in hadronic matter produced in central Au+Au collisions 
at RHIC energies $\sqrt {s_{NN}}=$ 130 and 200 GeV. In calculations the
nucleon and antinucleon distributions in hadronic matter are results of
evolution from their freeze-out distributions which well fit the experimental
$p_T$ spectra of proton and antiproton. Any measured
$\chi_{cJ}$ mesons at low $p_T$ are generated from deconfined matter and give
an explicit proof of regeneration mechanism (recombination mechanism).     
\end{abstract}

\renewcommand\@

Experimental measurements have revealed new phenomena in Au+Au collisions at 
the Relativistic Heavy Ion Collider (RHIC). Basing with 
the experimental results,
we derive new results in this letter. We predict that $\chi_{cJ}$ mesons
at low transverse momentum almost vanish 
in central Au+Au collisions at RHIC energies $\sqrt 
{s_{NN}}=$ 130 and 200 GeV because of the collisions from nucleons and 
antinucleons in hadronic matter.

Compared to the widely studied $J/\psi$ physics, $\chi_{cJ}$ physics in
nucleus-nucleus collisions is little known even at the CERN-SPS energies.  
However, the HERA-B  and NA60 collaborations have been measuring 
the $\chi_{cJ}$ suppression in proton-nucleus collisions \cite{ZA,SP}. 
In future the PHENIX collaboration will measure $\chi_{cJ}$ in their
upgrade programs \cite{DA}.
Since copious $\chi_{cJ}$ mesons are expected to be produced 
by the initial nucleus-nucleus collisions at RHIC energies, and the measured 
baryon/meson ratios increase with transverse momentum in the central Au+Au
collisions \cite{AK1,ASS1}, it is important
to study $\chi_{cJ}$ survival probability in connection with the collisions
with nucleons and antinucleons.
 
At present the transverse mass spectra of light hadrons have been measured 
\cite{AK1,ASS1,AKH,LRA,AK2} and 
can be accounted for by the hydrodynamic model \cite{HU,HT,XXM1,HP,MK,TD,EKJ}. 
Since we attempt to calculate $\chi_{cJ}$ survival probability in connection 
to the nucleon and antinucleon distributions in hadronic matter, the 
data measured by PHENIX collaboration are applied to get freeze-out 
distributions. In Ref. \cite{XXM1}, the STAR collaboration data \cite{AKH}
was used, but the hadron spectra were measured at
the transverse momentum $p_T<1$
GeV/$c$ at $\sqrt {s_{NN}}=130$ GeV. 
Therefore, we have to renew the freeze-out distributions of proton and
antiproton obtained in Ref. \cite{XXM1} to fit the PHENIX collaboration data  
up to $p_T=4.5$ GeV/$c$ at $\sqrt {s_{NN}}=130$ and 200 GeV \cite{AK1,ASS1}.

Assume that the Boltzmann form is satisfied by
baryon and antibaryon distributions at a given time $\tau$
in hadronic matter with temperature $T$:
\begin{equation}
f(\lambda,k,T)=g\lambda(\tau,\eta,r)e^{-k\cdot u/T}
\end{equation}
with the degeneracy factor $g=2$, fugacity $\lambda$, 
transverse radius $r$,
space-time rapidity $\eta$, the four-velocity $u$ of fluid flow and the 
baryon four-momentum $k$  in the center-of-mass frame of hadronic matter, 
\begin{equation}
k=(m_{\mathrm{kT}}\cosh y, k_{\mathrm{T}}\cos\varphi, 
k_{\mathrm{T}}\cos\varphi, m_{\mathrm{kT}}\sinh y),
\end{equation}
where $m_{\mathrm{kT}}$, $k_{\mathrm{T}}$ and $y$ 
are the transverse mass, transverse momentum and rapidity,
respectively, and $\varphi$ is the angle between 
the transverse momentum of hadron and the
transverse momentum  of $J/\psi$. 
Derived via the Lorentz 
transformation, the baryon energy observed in a flow cell is
\begin{equation}
k\cdot u=[m_{\mathrm{kT}}\cosh (y-\eta)-v_rk_{\mathrm{T}}\cos(\varphi-\phi)]/
\sqrt{1-v_r^2}
\end{equation}
where $v_r$ is the transverse velocity of flow cell and $\phi$ is the angle
between $\vec {v}_r$ and the transverse momentum of $J/\psi$.

The Lorentz-invariant spectra at the freeze-out time $\tau_{\mathrm{fh}}$ 
in the central rapidity region is obtained
from the Cooper-Frye formula \cite{CF}:
\begin{eqnarray}
\frac{d^2N(\tau_{\mathrm{fh}})}{k_{\mathrm{T}}dk_{\mathrm{T}}dy} 
& =& \frac{g\tau_{\mathrm{fh}}}
{(2\pi)^2}\int_0^{R(\tau_{\mathrm{fh}})}dr
\int_{\eta_{\mathrm{min}}}^{\eta_{\mathrm{max}}}d\eta
\int_0^{2\pi}d\phi r\lambda_{\mathrm{fh}}m_{\mathrm{kT}}\cosh(-\eta)\nonumber\\
& & \exp \{ -[m_{\mathrm{kT}}\cosh(-\eta)-v_rk_{\mathrm{T}}\cos\phi]
/T_{\mathrm{fh}}\sqrt{1-v_r^2} \},
\end{eqnarray}
where hadronic matter  freezes out with
the transverse radius $R(\tau_{\mathrm{fh}})$ and
the space-time rapidity spans from $\eta_{\rm min}=-5.5$ 
to $\eta_{\rm max}=5.5$.

We get two sets of fugacities and transverse velocities. The first set is
\begin{equation}
\lambda_{\mathrm{fh}}\propto\left\{
\begin{array}{ll}
1,&\quad r<a \\
e^{-(r-a)^2},&\quad r\ge a
\end{array}
\right.
\begin{array}{c}
v_r=r/14.8
\end{array}
\end{equation}
which shows a linear dependence of transverse velocity on $r$ and the second
set is
\begin{equation}
\lambda_{\mathrm{fh}}\propto\left\{
\begin{array}{ll}
1,&\quad r<b\pi \\
e^{-(r/b-\pi)},&\quad r\ge b\pi
\end{array}
\right.
\begin{array}{c}
v_r=\tanh(r/11.6).
\end{array}
\end{equation}
which exhibits a hyperbolic dependence of transverse velocity on $r$.
In each set the fugacities of proton and antiproton have the same form but a
constant normalized to the experimental data.

The parameters a and b are determined in the fit of our 
theoretical $p_T$ spectra
at midrapidity to the experimental data  for the central Au+Au
collisions at $\sqrt {s_{NN}}=130$ GeV \cite{AK1} and 200 GeV \cite{ASS1}. 
Results are listed in Table 1 along with the proper time
$\tau_{\mathrm{f}}$ at which the parton plasma hadronizes and 
$\tau_{\mathrm{fh}}$ at which hadronic matter freezes out 
with the temperature $T_{\mathrm{fh}}=0.13$ GeV.

\begin{table}[h] 
\caption {Parameters for the central Au+Au collisions}
\begin{center}
\begin{tabular}{|c|ccccc|} \hline
$\sqrt{s_{NN}}$(GeV) & $a$ & $b$ & $\tau_{\mathrm{f}}$(fm/c) 
& $\tau_{\mathrm{fh}}$(fm/c) 
& $R(\tau_{\mathrm{fh}})$(fm) \\
\hline
130 & 9 & 2 & 4 & 9.6 & 12.2 \\
200 & 10 & 2.3 & 4.3 & 10.3 & 12.6 \\
\hline
\end{tabular}
\end{center}
\end{table}

Here the values of the hadronization time $\tau_{\rm f}$ are on the scale of
4 fm/$c$ given in Ref. \cite{LP} and the freeze-out times are similar to that
provided in Ref. \cite{XXM2}. 
The two transverse velocities are independent of $\sqrt {s_{NN}}$.
The transverse momentum spectra obtained from Eq. (4)
are compared to the experimental data \cite{AK1,ASS1} in Figs. 1 and 2.     

\begin{figure}[ht]
  \centering
  \begin{minipage}[t]{0.85\textwidth}
    \centering
    \includegraphics[width=4.8 in]{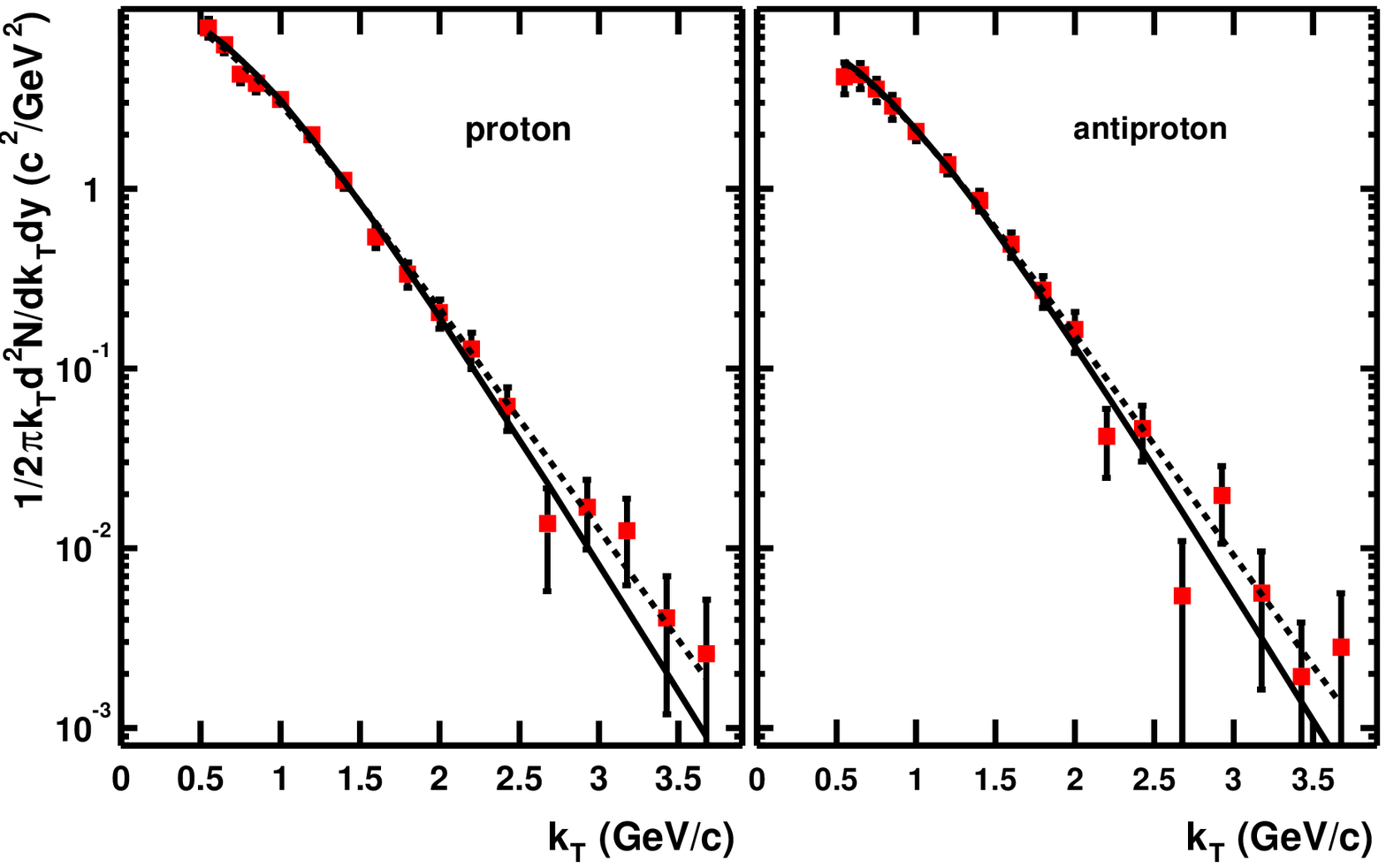}
    \caption{Proton and antiproton invariant yields 
in the central rapidity region
with $\lambda_{\mathrm{fh}}$ and $v_r$ in Eqs. (5) (solid) and (6) 
(dashed) at $\sqrt{s_{NN}}=130$ GeV.}
  \end{minipage}\\[20pt]
  \centering
  \begin{minipage}[t]{0.85\textwidth}
    \centering
    \includegraphics[width=4.8 in]{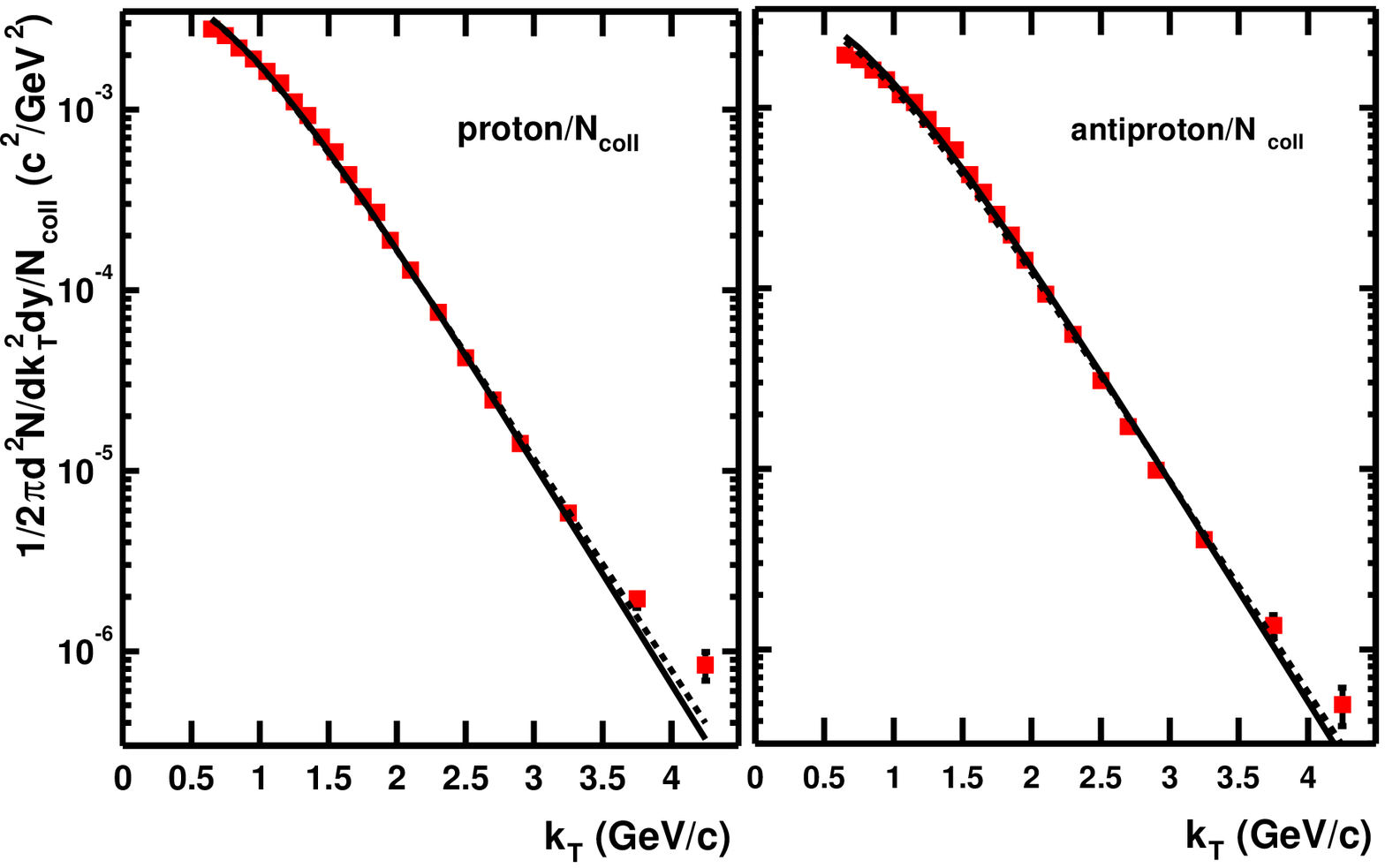}
    \caption{Proton and antiproton invariant yields scaled by $N_{\rm coll}$,
which is the average number of binary nucleon-nucleon collisions,  
in the central rapidity region with $\lambda_{\mathrm{fh}}$ 
and $v_r$ in Eqs. (5) (solid) and (6) (dashed) 
at $\sqrt{s_{NN}}=200$ GeV.}
  \end{minipage}
\end{figure}

With either of the two sets of 
the fugacities and the transverse velocities at freeze-out, we can obtain 
$v_r(\tau,r)$ at any position and any time prior to freeze-out by the
hydrodynamic equation regarding the energy-momentum tensor 
and $\lambda(\tau,r)$ by the number conservation 
followed by the evolution equation \cite{SDK}:
\begin{equation}
\partial_t\lambda+v_r\partial_r\lambda+\frac{\lambda}{\gamma T^3}
\partial_t(\gamma T^3)
+\frac{\lambda v_r}{\gamma T^3}\partial_r(\gamma T^3)+\lambda\partial_rv_r
+\lambda(\frac{v_r}{r}
+\frac{1}{t})=0
\end{equation}  
where $\gamma=1/\sqrt{1-v_r^2}$. The evolution equation is numerically solved
until the temperature of hadronic matter very close to the phase 
transition point of lattice gauge results $T_c=0.175$ GeV \cite{KF}.  

Up to now the distribution functions of protons and antiprotons 
have been obtained. Then we start studying the interactions of $\chi_{cJ}$ with
nucleons and antinucleons in  hadronic matter. 
  The thermal average of the product of the
dissociation cross section $\sigma$ and the relative velocity
$v_{\rm rel}$ of baryon and $\chi_{cJ}$ is 
\begin{equation}
\langle\sigma(s)v_{\mathrm{rel}}\rangle=\frac{1}{n(\tau)}
\int \frac {d^3k}{(2\pi)^3}\sigma(s)v_{\mathrm{rel}}f(\lambda, k, T)
\end{equation}
where $n(\tau)$ is the baryon number density 
\begin{equation}
n(\tau)=\int \frac{d^3k}{(2\pi)^3} f(\lambda, k, T)
\end{equation}
The baryon-$\chi_{cJ}$ dissociation cross 
section which depends on the center-of-mass energy of baryon 
and $\chi_{cJ}$ is taken from Ref. \cite{XXM3}.
 
We consider that a $\chi_{cJ}$ with transverse momentum $P_T$ at midrapidity
is produced at the position $r'$ while the two colliding gold nuclei completely
overlap. The initial $\chi_{cJ}$ production is assumed to be proportional 
to the number of binary nucleon-nucleon 
interactions $N_A(r')=A^2(1-r'^2/R_A^2)/2\pi R_A^2$. 
Integrating $\vec {r}'$, we obtain the survival probability 
of initially produced $\chi_{cJ}$ in hadronic matter:
\begin{equation}
S(P_T)=\frac{\int d^2r'(R_A^2-r'^2)
\exp[-\int_{\tau_{\mathrm{f}}}^{\tau_{\rm min}}d\tau 
n(\tau)\langle\sigma(s)v_{\mathrm{rel}}\rangle]}{\int d^2r'(R_A^2-r'^2)}
\end{equation}  
where  $\tau_{\rm min}$ is the smaller one of the 
freeze-out time and $\tau_{\mathrm{max}}$
the maximum time of stay of $\chi_{cJ}$ in hadronic matter, 
$\tau_{\rm min}=\mathrm{min}(\tau_{\mathrm{fh}},\tau_{\mathrm{max}})$. 
The time $\tau_{\mathrm{max}}$ depends on the $\chi_{cJ}$ velocity and the
flow velocity of lateral face of hadronic matter $v_{\rm sf}$ by
\begin{eqnarray}
\tau_{\rm max} & = & \frac {M_TP_T}{M_T^2v_{\rm sf}^2 -P_T^2}   
 \bigg ( r' \cos \varphi' -\frac {R_AM_Tv_{\rm sf}}{P_T}   \cr
& & -\sqrt {R_A^2 +v_{\rm sf}^2r'^2\frac {M_T^2}{P_T^2}
-2R_Av_{\rm sf}r'\cos \varphi' \frac {M_T}{P_T} -r'^2 \sin^2 \varphi'} 
\bigg )    
\end{eqnarray}
where $M_T$ is the $\chi_{cJ}$ transverse mass,
$R_A$ is the radius of glod nucleus and
 $\varphi'$ is the angle between the position vector $\vec {r}'$ and 
$\vec {P}_T$. We count time from the moment
when the two colliding nuclei completely overlap.
 
Since the neutron and antineutron yields have not been measured, 
we assume that neutron and antineutron have the same distribution functions 
as proton and antiproton, respectively.  The $\chi_{cJ}$ dissociation cross
section in collision with an antiproton, a neutron or an antineutron is 
identical to the cross section with a proton. The survival probability as a
function of $P_T$ at midrapidity is calculated and plotted in Fig. 3 
where the left and right panels show results of using the freeze-out
fugacities and transverse velocities in Eqs. (5) and (6), respectively.   

\begin{figure}[ht]
  \centering
  \begin{minipage}[t]{0.85\textwidth}
    \centering
    \includegraphics[width=4.8 in]{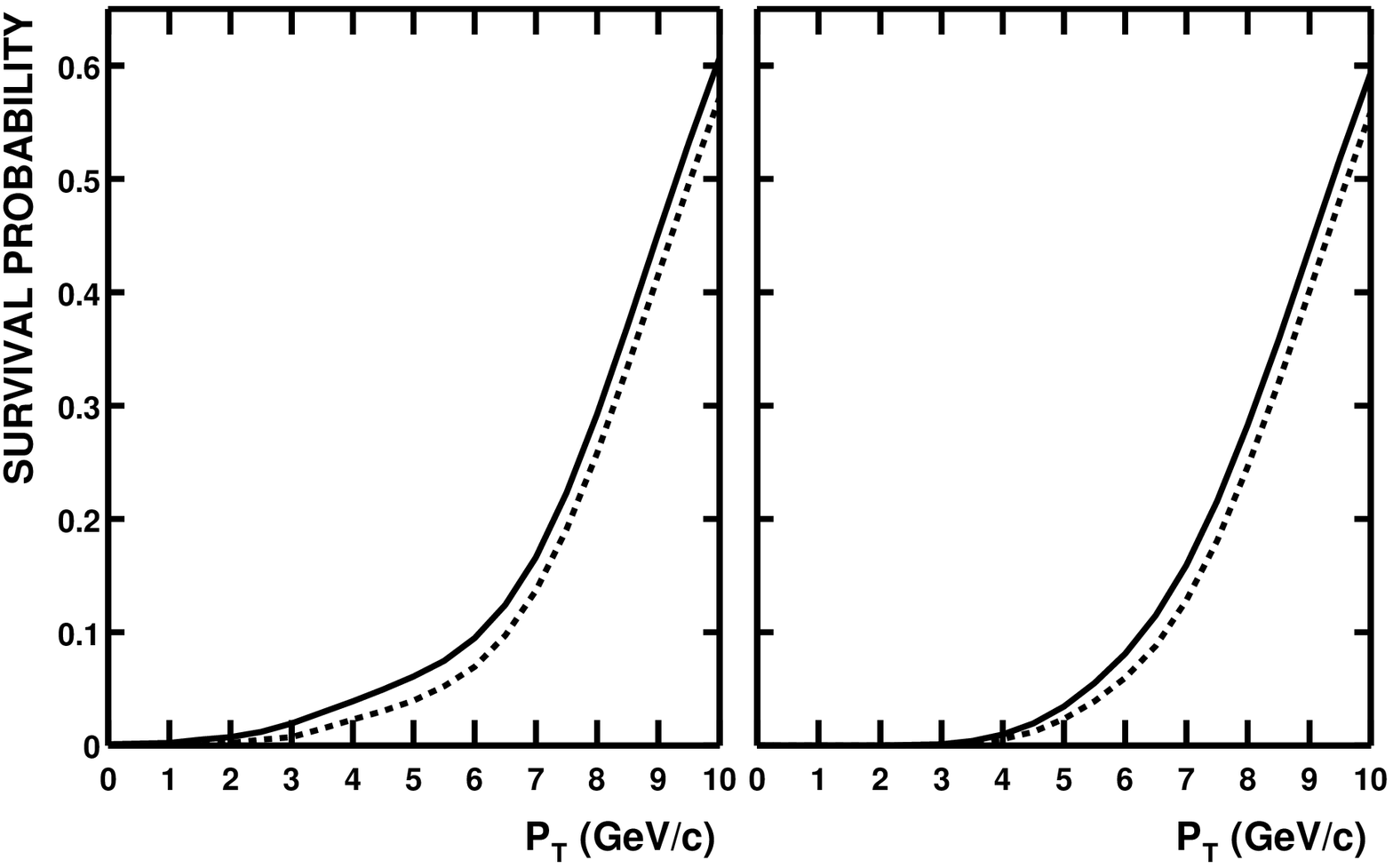}
    \caption{Survival probability of $\chi_{cJ}$ in collisions with 
nucleons and antinucleons in hadronic matter. Solid curves: 
$\sqrt{s_{NN}}=130$ GeV; Dashed curves: $\sqrt{s_{NN}}=200$ GeV. 
Left panel: Eqs. (5); Right panel: Eqs. (6).} 
  \end{minipage}
\end{figure}    
  
The survival probability reflects the extent to which a $\chi_{cJ}$ meson 
survives after the collisions with nucleons and antinucleons in hadronic 
matter. In either of the two panels, the dashed curve for the central Au+Au
collision at $\sqrt{s_{NN}}=200$ GeV is close to the solid curve at
$\sqrt{s_{NN}}=130$ GeV, but slightly lower because the nucleons and 
antinucleons are denser in the higher-energy Au+Au collision.  
A larger transverse momentum  
corresponds to a larger center-of-mass energy of $\chi_{cJ}$ 
and nucleon (antinucleon), 
which leads to a smaller $\chi_{cJ}$ dissociation cross section \cite{XXM3}. 
At large $P_T$, more
$\chi_{cJ}$  can escape quickly from hadronic matter. The two factors make 
the survival probability increasing rapidly when the transverse momentum 
of $\chi_{cJ}$ is larger than 6 GeV/c. At low $P_T$, the 
$\chi_{cJ}$ velocity doesn't exceed the   
flow velocity of lateral face of hadronic matter, $v_{\rm sf}=0.6c$.
Hence $\chi_{cJ}$ stays inside hadronic matter until freeze-out.
The survival probability at low $P_T$ approaches zero as shown by Fig. 3. 
and $\chi_{cJ}$ mesons are almost suppressed by the nucleons and
antinucleons alone. This result is meaningful to the coming 
RHIC experiments as explained below.

It is known that experimentalists measure prompt $J/\psi$ production which 
includes the direct $J/\psi$ production, the radiative feeddown from 
$\chi_{cJ}$ and the decay of $\psi'$.  
The contribution to the prompt $J/\psi$ production from 
$\chi_{cJ}$ occupies about 30\% \cite{RL}. On the other hand,
the measurements of the PHENIX collaboration have exhibited enhanced
baryon production compared to the suppressed pion production in the central
Au+Au collisions at $\sqrt{s_{NN}}=130$ and 200 GeV  \cite{AK1,ASS1}. 
Therefore, the result of the very strong
suppression of initial $\chi_{cJ}$ will affect the prompt $J/\psi$ production. 

The nucleus-nucleus collisions at RHIC 
and LHC energies have been divided into four stages: 
initial parton-parton scatterings, prethermal parton matter, 
thermalized parton plasma and hadronic matter \cite{BTS}. 
Study of Ref. \cite{XXM2} suggested that the prompt $\chi_{cJ}$ production 
around $P_T=3$ GeV at midrapidity
may be enhanced by the yield of color singlet and color octet $c\bar c$ 
pairs from
the partonic system in the prethermal and thermal stages. If the yield can 
compete with the amount of $\chi_{cJ}$ dissociated, the survival probability
of $\chi_{cJ}$ cannot be zero at low $P_T$. An observation offering finite 
survival probability would hint the existence of
deconfined matter in Au+Au collisions at RHIC energies. 
In the first measurement of $J/\psi$ production in Au+Au collisions 
a good analysis on the central collision \cite{ASS2} was not conducted.
But we can speculate two cases for the central collision: 
A sudden drop
of data in the central collision would reflect a strong suppression of
$J/\psi$, $\chi_{cJ}$ and $\psi'$; A flat or even rising transition of data
from collisions at midcentrality to the central collision would indicate a 
large amount of prompt $J/\psi$ yielded from the deconfined matter. Therefore,
any experimental result in future will give a feedback to examine our 
prediction of the very strong suppression 
of initial $\chi_{cJ}$ in the central collision.

Recent lattice gauge
results give a critical temperature of 0.151 GeV \cite{AFKS}. If this value is 
used, the lifetime of hadronic matter gets shorter but any initially 
produced $\chi_{cJ}$ at low $p_T$ are also almost
dissociated since gluons, quarks and antiquarks in quark-gluon plasma
can break $\chi_{cJ}$. Therefore, any
measured $\chi_{cJ}$ mesons at low $p_T$ at midrapidity
offer a proof of regeneration mechanism (recombination mechanism) \cite{REGC}. 
  
In summary, we have calculated survival probability of initially produced
$\chi_{cJ}$ at midrapidity
with the nucleon and antinucleon distributions in the central Au+Au collisions
at  $\sqrt {s_{NN}}=130$ and 200 GeV.  The survival probability 
is only related to the collisions of $\chi_{cJ}$ and nucleons and antinucleons.
The prominent result is that the survival probability at low $P_T$ 
is almost zero, which significantly affect the $J/\psi$ 
suppression at low $P_T$, but any measured $\chi_{cJ}$ at low $P_T$ 
must come from deconfined matter and   
obviously exhibits regeneration mechanism (recombination mechanism).

\end{document}